\documentclass[12pt,thmsa]{article}%
\usepackage{sw20elba}
\usepackage{amsmath}
\usepackage{graphicx}%
\usepackage{amsfonts}%
\usepackage{amssymb}

\begin{document}

\begin{center}
\textbf{Dependence of the Superconducting Transition Temperature of Single-
and Polycrystalline MgB}$_{2}$\textbf{ Hydrostatic Pressure}
\end{center}

\bigskip

\begin{center}
S. Deemyad,$^{a}$ T. Tomita,$^{a}$ J. J. Hamlin,$^{a}$ B.R. Beckett,$^{a}$ J.
S. Schilling,$^{a}$ D. G. Hinks,$^{b}$ J. D. Jorgensen,$^{b}$ S. Lee,$^{c}$ S.
Tajima$^{c}$\bigskip

$^{a}$\textit{Department of Physics, Washington University, CB 1105, One
Brookings Dr., St. Louis, MO 63130, USA}

$^{b}$\textit{Materials Science Division, Argonne National Laboratory, 9700
South Cass Avenue, Argonne, IL 60439, USA}

$^{c}$\textit{Superconductivity Research Laboratory, ISTEC, 1-10-13 Shinonome,
Koto-ku, Tokyo 135-0062, Japan}

\bigskip
\end{center}

\noindent\textbf{Abstract. \ }The dependence of $T_{c}$ for MgB$_{2}$ on
purely hydrostatic or nearly hydrostatic pressure has been determined to 23
GPa for single-crystalline and to 32 GPa for polycrystalline samples, and
found to be in good agreement. $T_{c}$ decreases from 39 K at ambient pressure
to 15 K at 32 GPa with an initial slope $dT_{c}/dP\simeq$ -1.11(2) K/GPa.
Evidence is presented that the differing values of $dT_{c}/dP$ reported in the
literature result primarily from shear-stress effects in nonhydrostatic
pressure media and not differences in the samples. Although comparison of
these results with theory supports phonon-mediated superconductivity, a
critical test of theory must await volume-dependent calculations based on the
solution of the anisotropic Eliashberg equations. \bigskip

\section{Introduction}

In spite of considerable efforts during the 20 months since the discovery of
superconductivity in MgB$_{2}$ at 39 K \cite{r0}, the search for a further
binary compound with a higher value of $T_{c}$ has yet to bear fruit
\cite{r1}. \ In contrast, the high-$T_{c}$ oxide Y-Ba-Cu-O ($T_{c}\approx$ 92
K) was synthesized \cite{r2} only a few weeks after the landmark discovery of
superconductivity near 35 K in La-Ba-Cu-O \cite{r3}. \ Progress with the
binary compounds may have to wait until we first reach a clear understanding
of MgB$_{2}$'s extraordinary superconducting and normal-state properties, most
of which are highly anisotropic. \ The compressibility, for example, is 64\%
larger along the $c$ axis than along the $a$ direction \cite{r22}. \ Under
$c$-axis compression the electronic density of states $N(E_{f})$ is predicted
to decrease much more rapidly ($\sim$ 0.6\%/GPa) than under hydrostatic
compression ($\sim$ 0.1\%/GPa) \cite{kobayashi1}, the $\sigma$ band shifting
downward and the hole number in the $\sigma$ band\ decreasing \cite{r24}.
\ The anisotropy in the upper critical field\ $H_{c2}$ increases strongly with
decreasing temperature \cite{hc21,hc22}, approaching a value near 7 at 0 K
\cite{hc21}. \ A large number of experiments point to the existence of at
least two superconducting energy gaps \cite{gaps} which are predicted
\cite{theory1,theory2,r4} to open up on different parts of the anisotropic
Fermi surface. \ Gonnelli \textit{et al.}\cite{gonnelli} have very recently
provided direct evidence for two-band superconductivity in MgB$_{2}$ where the
temperature-dependence of the distinct gaps originating from the $\sigma$ and
$\pi$ bands were measured using point-contact spectroscopy on single crystals.

The existence of multiple gaps and the highly anisotropic electronic and
lattice-vibration properties of MgB$_{2}$ have recently prompted Choi
\textit{et al.} \cite{r4} to attempt the solution of the fully anisotropic
Eliashberg equations. \ A comparison of their calculation with experiment has
yielded promising results \cite{theory2,r4}. \ A critical test of their
approach would be provided by extending their calculation to reduced lattice
parameters, allowing a direct comparison with the results of high pressure experiments.

Following the discovery of a new superconductor, high pressure studies are
often among the first to be carried out. \ The reason for this is that the
magnitude and sign of $dT_{c}/dP$ help guide the materials scientist how to
best modify the superconductor to raise $T_{c}$ at ambient pressure, a case in
point being the discovery of Y-Ba-Cu-O \cite{r2}. \ In addition, the
high-pressure technique, sometimes in conjunction with high temperatures, is
also invaluable to: \ (1) create new superconductors (e.g. sulfur
\cite{sulfur} or oxygen \cite{oxygen} which metallize at Mbar pressures); (2)
vary the properties of known superconductors (e.g. polycrystalline MgB$_{2}$
\cite{hpht} or the synthesis of MgB$_{2}$ single crystals \cite{lee,r5,jung});
(3) induce structural phase transitions (e.g. in Ba, As, Bi, Sn, Ga and Tl
\cite{brand}); and, finally, (4) vary the lattice parameters to help identify
the pairing mechanism and critically test theoretical models. \ Unfortunately,
this final application has been traditionally underutilized by theorists.

Hydrostatic or uniaxial pressure experiments on single crystals determine the
dependence of a given property solely on the lattice parameters. \ \ Most
high-pressure experiments, however, are carried out under nonhydrostatic
conditions which subject the sample to a distribution of unspecified shear
stresses. \ These shear stresses may be large enough to plastically deform the
sample, resulting in lasting distortions and lattice defects. \ The pressure
dependence of $T_{c}$ may thus depend on whether the pressure medium is
hydrostatic or not, particularly in elastically anisotropic materials like
quasi 1D and 2D organic superconductors \cite{jerome} or high-$T_{c}$ oxides
\cite{sp}. \ Studies on single crystals are of particular value since strain
effects from grain boundaries in polycrystalline materials are avoided.

Several studies of the dependence of $T_{c}$ on pressure for polycrystalline
MgB$_{2}$ were carried out shortly after the discovery of its
superconductivity \cite{r12,r13,r14,r15}. \ Using solid steatite pressure
medium to 20 GPa, Monteverde \textit{et al.} \cite{r12} reported that $T_{c}$
decreased under pressure at different initial rates (-0.35 to -0.8 K/GPa) for
each of the four samples studied. \ On the other hand, in an experiment in
fluid Fluorinert to $\sim$ 1.7 GPa, Lorenz \textit{et al.} \cite{r13} and
Saito \textit{et al.} \cite{r14} found $dT_{c}/dP\simeq$ -1.6 K/GPa and -1.9
K/GPa, respectively. \ The first truely hydrostatic measurement of $T_{c}(P)$
was carried out by our group to 0.7 GPa using He gas on an isotopically pure
($^{11}$B) sample, revealing that $T_{c}$ decreases reversibly under
hydrostatic pressure at the rate $dT_{c}/dP\simeq$ -1.11(2) K/GPa \cite{r15};
\ later work on isotopically pure $^{10}$B and $^{11}$B samples yielded
dependences between -1.09(4) K/GPa and -1.12(3) K/GPa \cite{r17}. \ In further
He-gas studies, Lorenz \textit{et al.} \cite{r18} and Schlachter \textit{et
al.} \cite{r19} obtained -1.07 K/GPa and -1.13 K/GPa, respectively.
\ Experiments utilizing fluid pressure media were carried out by Razavi
\textit{et al.} (-1.18(6) K/GPa) \cite{razavi}, Choi \textit{et al.} (-1.36
K/GPa) \cite{choi}, and Kazakov \textit{et al.} (-1.5 K/GPa) \cite{kazakov}.
\ The first high-pressure measurements on a single crystal were carried out by
Masui \textit{et al.} \cite{masui} using Fluorinert, yielding $dT_{c}%
/dP\simeq$ -2.0 K/GPa. \ In the above experiments the values of $dT_{c}/dP$
reported are seen to differ by more than a factor of two. \ Experiments to
much higher pressures have been carried out by several groups and will be
discussed below. \ The results of all known $dT_{c}/dP$ measurements on
MgB$_{2}$ are summarized in the Table.

In this paper we provide evidence that the variation in the reported
$T_{c}(P)$ dependences for MgB$_{2}$ is primarily a result of shear stresses
exerted by the solidified pressure media on the sample. \ From recent
measurements on high quality single-crystalline and polycrystalline samples we
conclude that the initial dependence of the transition temperature on purely
hydrostatic pressure is given by $dT_{c}/dP\simeq-1.11(2)$ K/GPa \ \ Under
nearly hydrostatic (dense He) pressures to 32 GPa ($V/V_{o}\simeq$ 0.855), we
find $T_{c}$ to decrease monotonically from $\sim$ 39 K to 15 K. \ Although
these results appear consistent with phonon-mediated superconductivity in
MgB$_{2},$ a quantitative check must await comparison with calculations based
on the fully anisotropic Eliashberg equations.

\section{Experimental methods}

A wide variety of high-pressure techniques with many different pressure media
have been used to study the dependence of $T_{c}$ on pressure \cite{eremets}.
\ All techniques discussed below have been used at some time by our group.

Perhaps the most widely used technique in the pressure range 1-2 GPa is a
piston-cylinder cell in which two pistons compress a Teflon bucket containing
the sample immersed in a fluid pressure medium such as Fluorinert,
n-pentane/isopentane, or silicon oil. \ For studies to higher pressure a
diamond-anvil-cell (DAC) may be used with a 4:1 methanol:ethanol mixture as
pressure medium since this mixture remains fluid to approximately 10 GPa at
ambient temperature. \ All these fluid pressure media freeze upon cooling and
subject the immersed sample to shear stresses of varying strength which depend
on the pressure medium used, the details of the pressure technique, and the
rate at which the pressure cell is cooled, among other factors.

The only pressure medium that remains fluid under high pressure ($P\leq$ 0.4
GPa) near 40 K, where MgB$_{2}$ superconducts, is helium and thus only
high-pressure experiments on MgB$_{2}$ in liquid helium are able to determine
the dependence of $T_{c}$ on purely hydrostatic pressure. \ At pressures above
0.5 GPa, helium is frozen below 39 K, but the pressure is still very nearly
hydrostatic since solid helium is the softest solid known; in addition, if the
proper cooling procedure is followed, a single crystal of helium can be grown
around the sample, resulting in negligible shear stresses which permit even de
Haas van Alphen measurements on single crystals at very low temperatures
\cite{r9}. \ High-pressure studies in liquid helium are thus the
measurements-of-choice for exacting quantitative studies of the properties of
solids under high-pressure conditions.

At the other end of the spectrum, in some high pressure techniques solid
pressure media (e.g.steatite, NaCl or NaF) are used \cite{eremets}. \ With a
solid pressure medium, the application of pressure may subject the sample to
relatively large shear stresses, possibly strong enough to plastically deform
the sample or even crush it, thus introducing a large number of lattice
defects. \ The magnitude and direction of the shear stresses depend, among
other things, on the pressure medium used, the temperature at which the
pressure is changed, the pressure range, and whether the ring containing the
solid pressure medium is supported by a ``belt'' or not. \ With this so-called
``quasihydrostatic'' technique $T_{c}$ may not be a reversible function of
pressure. \ For these reasons high-pressure techniques using solid pressure
media should be avoided in quantitative studies, particularly when studying
elastically anisotropic materials such as the high-$T_{c}$ oxides \cite{sp},
organic superconductors \cite{jerome}, and MgB$_{2}.$

In the present experiments, dense helium is used as pressure medium both in a
He-gas pressure cell to 1 GPa and in a DAC to over 30 GPa. \ The pressure in
the cell is measured at temperatures within a few degrees of the transition
temperature $T_{c};$ the superconducting transition is detected in a sensitive
ac susceptibility measurement \ Details of these pressure techniques have been
published elsewhere \cite{r11}. \ Isotopically pure polycrystalline MgB$_{2}$
samples were synthesized at the Argonne National Labs \cite{r22,hinks}.
\ Single crystals were grown at the Superconductivity Research Laboratory
(ISTEC) in Tokyo, Japan; following the crystal growth at elevated
temperatures, ``type A'' (``type B'') crystals were quenched (slow cooled)
\cite{lee}. \ The dimensions of the crystals studied in the He-gas cell were
approximately $0.15\times0.4\times0.03$ mm$^{3}$ and in the DAC $\sim
0.09\times0.09\times0.03$ mm$^{3}.$

\section{Results of experiment}

In Figs. 1-3 we show the results of our measurements of the pressure
dependence of $T_{c}$ on MgB$_{2}$ single crystals. \ The midpoint of the
superconducting transition in the real part of the ac susceptibility
$\chi^{\prime}$ is used to define $T_{c}$ \cite{midpoint}, as seen in Fig. 1.
\ The two ``type B'' crystals have higher values of $T_{c}$ ($\sim$ 38.2 K)
than the ``type A'' crystal ($\sim$ 37.2 K). \ For both ``type A'' and ``type
B'' crystals $T_{c}$ is seen in Fig. 2 to decrease reversibly and linearly
with hydrostatic (He-gas) pressure. \ As with the polycrystalline samples, the
same $T_{c}(P)$ dependence is obtained whether the pressure is changed at low
temperatures or at room temperature. \ As seen in the Table, the pressure
dependence of $T_{c}$ in the He-gas experiments is very nearly the same
whether MgB$_{2}$ is in single- or polycrystalline form.

In Fig. 3 we extend the He-gas results on the same ``type B'' single crystal
\cite{crystal} to much higher pressures using a He-loaded DAC. \ The initial
$T_{c}(P)$ dependence agrees well with the data in Fig. 2, showing a positive
curvature at higher pressures. \ As seen in Fig. 3, this curvature is removed
if $T_{c}$ is plotted versus the relative unit-cell volume $V/V_{o}$
\cite{murg}. \ The linearity of the $T_{c}$ versus $V/V_{o}$ data over such a
wide range of pressure is remarkable.

In Fig. 4 are shown our $T_{c}(P)$ data on an isotopically pure $^{11}$B
polycrystalline sample to 32 GPa from three separate experiments in a
He-loaded DAC \cite{r21}; the sample used is from the same synthesis batch as
in our earlier He-gas studies to 0.7 GPa \cite{r15}. \ $T_{c}$ decreases
monotonically and reversibly with pressure from $\sim$ 39 K at ambient
pressure to $\sim$ 15 K at 32 GPa. \ The initial slope, $dT_{c}/dP\simeq$ -1.1
K/GPa, is the same as in the He-gas data. \ As will be discussed below (see
Fig. 7), the single-crystal and polycrystalline data are in good agreement.

The only other $T_{c}(P)$ measurements on polycrystalline MgB$_{2}$ to very
high pressures using dense He as pressure medium were carried out by Struzhkin
\textit{et al.} \cite{r20} and are included in Fig. 4 for the $^{11}$B
isotope; these authors use a double-modulation ac susceptibility technique
which determines the superconducting onset rather than the superconducting
midpoint. \ As seen in Fig. 4, the agreement with our data is remarkably good
to 20 GPa, but begins to deviate at higher pressures. \ Parallel studies by
the same authors \cite{r20} to 44 GPa in dense He on a $^{10}$B isotopic
sample yield a similar $T_{c}(P)$ dependence which lies $\sim$ 1 K above their
$^{11}$B data below 20 GPa, but gradually merges at higher pressures. \ Unlike
our data in Figs. 3 and 4, where $T_{c}$ is seen to be a linear function of
$V/V_{o},$ the data of Struzhkin \textit{et al.} \cite{r20} show a break in
slope $dT_{c}/dV$ near 15-20 GPa. \ For a full discussion of the latter data
see the paper by Goncharov and Struzhkin in this special edition.

Non-He-gas studies to very high pressures include those of Tang \textit{et
al.} \cite{tang} to 9 GPa using a cubic-anvil cell with Fluorinert pressure
medium; the $T_{c}(P)$ dependence is in good agreement with that in Fig. 4.
\ However, in DAC studies using methanol-ethanol, Tissen \textit{et al.}
\cite{r6} found a much larger initial slope ($dT_{c}/dP\approx$ -2 K/GPa)
accompanied by a relatively large drop in $T_{c}$ to approximately 6 K at 28
GPa; these authors interpret a break in slope $dT_{c}/dP$ near 10 GPa as
evidence for a topological transition. \ We note that the present $T_{c}(P)$
data lie 1-2 K below those of Razavi \textit{et al.} \cite{razavi} to 11 GPa
obtained using steatite pressure medium. \ In further DAC measurements using
steatite Bordet \textit{et al.} \cite{bordet} and Monteverde \textit{et al.
}\cite{r12} report $T_{c}(P)$ dependences for four different samples which
generally lie well above those in the present measurement.

\section{Discussion}

\subsection{Intrinsic Dependence of T$_{c}$ on Pressure}

Before attempting a quantitative analysis of the present data, we would like
to first discuss possible origins for the differing values of $dT_{c}/dP$ for
MgB$_{2}$ reported in the literature (see the Table). \ Tissen \textit{et al.}
\cite{r6} and, somewhat later, Lorenz \textit{et al.} \cite{lorenz3} have
presented data indicating a strong inverse correlation between the magnitude
of the initial slope $dT_{c}/dP$ and the value of $T_{c}$ at ambient pressure,
i.e. $\left|  dT_{c}/dP\right|  _{0}$ is larger if $T_{c}(0)$ is smaller. \ To
reexamine this possible correlation, we have plotted in Fig. 5 the initial
slope $dT_{c}/dP$ using the high-pressure data in the Table. \ Taken as a
whole, the data in Fig. 5 would appear to give some support to the proposed
strong inverse correlation. \ Such a correlation is, however, \textit{not}
supported by the hydrostatic He-gas data. \ With the exception of the single
data point of Lorenz \textit{et al. }\cite{r18}, all known $(dT_{c}/dP)_{0}$
values for single- or polycrystalline samples obtained using He pressure
medium lie between -1.07 and -1.2 K/GPa. \ Although the He-gas data do not
support a strong correlation between $\left|  dT_{c}/dP\right|  _{0}$ and
$T_{c}(0)$, samples with lower values of $T_{c}(0) $ (37.16 K versus 39.1 K)
do appear to exhibit slightly ($\sim$ 5\%) larger initial slopes $dT_{c}/dP$
(-1.17 K/GPa versus -1.11 K/GPa). \ In contrast, the $(dT_{c}/dP)_{0}$ values
obtained using other less hydrostatic pressure media are often larger (up to
50-80\%) in magnitude. \ This indicates that differences in the values of
$(dT_{c}/dP)_{0}$ in the literature may depend more on the pressure medium
used than on differences between samples, at least as reflected in their
$T_{c}(0)$ values.

One way to help resolve this issue is to carry out parallel $T_{c}(P)$
measurements on the \textit{same} sample using \textit{different} pressure
media. \ We carried out two such studies. \ In the first, we placed a sample
provided by V. Tissen in our He-gas system, obtaining under purely hydrostatic
pressure conditions $dT_{c}/dP\simeq$ -1.2 K/GPa \cite{tissen2}. \ This value
is 40\% less than that (-2 K/GPa) from Tissen \textit{et al.'s} DAC
study\textit{\ }\cite{r6} on the \textit{same }sample with methanol/ethanol
pressure medium (see the appropriate vertical arrow in Fig. 5). \ In the
second experiment we took the same isotopically pure $^{11}$B sample studied
previously in both our He-gas and He-loaded DAC (where $dT_{c}/dP\simeq$ -1.1
K/GPa) and replaced the He pressure medium with Fluorinert, obtaining the DAC
data shown in Fig. 4 with initial slope $dT_{c}/dP\simeq$ -1.6 K/GPa, a nearly
50\% increase in slope (see the appropriate vertical arrow in Fig. 5).
\ Similar effects are observed with single crystals; the slope -2.0 K/GPa
observed in Fluorinert \cite{masui} is much larger than that (-1.1 K/GPa)
found in the present He-gas study on crystals from the same source. \ The
large values of $\left|  dT_{c}/dP\right|  _{0}$ reported in the literature
thus appear to arise from the use of fluid pressure media, such as Fluorinert
or methanol/ethanol, which freeze solid at temperatures well above
$T_{c}\approx$ 40 K. \ The present results thus do not support the existence
of the originally proposed \cite{r6,lorenz3} strong inverse correlation
between $\left|  dT_{c}/dP\right|  _{0}$ and $T_{c}(0).$

In is interesting to note that the DAC measurements of Schlachter \textit{et
al.} \cite{r19} with solid NaF pressure medium yield a large negative slope
with increasing pressure (-1.6 K/GPa) which reduces to -1.13 K/GPa as the
pressure is reduced to ambient, leaving an ambient pressure value $T_{c}(0)$
permanently suppressed by $\sim$ 12\% from the initial value. \ The authors
infer that, because of shear stress effects from the solid pressure medium,
the sample is degraded when pressure is applied, with no further degradation
upon pressure release. \ It is also noteworthy that the value $dT_{c}%
/dP\simeq$ -1.13 K/GPa is the same to 0.4 GPa as that obtained on the same
sample in a He-gas experiment by the same group (see Table).

With these results in mind, it is difficult to understand the relatively small
values of $\left|  dT_{c}/dP\right|  _{0}$ reported in the resistivity studies
by Monteverde \textit{et al.} \cite{r12} and Bordet \textit{et al.}%
\ \cite{bordet} using solid steatite pressure medium. \ Since their
temperature-dependent resistivity data at different pressures was not
published, it is difficult to speculate on possible origins for this
difference. \ The use by these authors of the resistivity onset\ to define
$T_{c}$ could lead to substantial errors in estimating $(dT_{c}/dP)_{0},$
particularly if significant broadening occurs in the transition under
pressure. \ We also note that in this type of pressure cell it is particularly
difficult to obtain reliable resistivity data and pressure values in the lower
1-2 GPa pressure range. \ In these papers \cite{r12,bordet} no information was
given whether $T_{c}(P)$ was reproducible for identical samples, reversible in
pressure, or whether the transition broadened significantly under pressure.
\ We note that Struzhkin \textit{et al.} \cite{r20} remeasured $T_{c}(P)$ in
their DAC with no pressure medium whatsoever, pressing the diamond anvils
directly onto the sample and gasket, and reported results resembling the two
lower curves of Monteverde \textit{et al.} \cite{r12}. \ It is also noteworthy
that\ Razavi \textit{et al.} \cite{razavi} also used steatite pressure medium,
and obtained the initial slope -1.03 K/GPa; in their published resistivity
data there is little change in the transition width with pressure so that the
estimate of the shift in $T_{c}$ is relatively straightforward.

In a further experiment, we left out the pressure medium entirely and pressed
two WC Bridgman anvils with 6 mm dia. flats for 5 minutes directly onto $\sim$
5 mg MgB$_{2}$ powder with 10 tons force; this typically results in a pressure
distribution across the center of the disc resembling a bell-shaped curve with
$P\approx$ 8 GPa at the center and $P=$ 0 at the edge. \ The sample was
thoroughly compacted by this procedure, resulting in a disc-shaped sample
approximately 70 $\mu$m thick. \ After removing the compacted disc from
between the anvils, the disc was separated into three separate regions
(center, middle, and outside), as seen in Fig. 6; the sample from each region
was then gently broken up and placed into our ac susceptibility coil system.
\ As seen in Fig. 6, the sharp superconducting transition of the virgin powder
sample is broadened by a significant amount ($\sim$ 10 K) by the compaction
procedure, the broadening being somewhat less from the center to the middle to
the outside. \ This broadening is presumably the result of strong plastic
deformation and the resulting internal strains in the material. \ If strong
internal strains can lead to broadening as large as 10 K at ambient pressure,
it is not unreasonable to assume that such strains are capable of causing
enhanced downward shifts in $T_{c}$ under nonhydrostatic pressure conditions.

The important role that defects and strains play in MgB$_{2}$'s
superconducting state is emphasized by the fact that the value of $T_{c}$ in a
large number of thin-film and bulk MgB$_{2}$ samples appears to follow a
Testardi correlation \cite{r1,lorenz3}: \ $T_{c}$ is lower for samples in
which the conduction electrons are strongly scattered. \ Lorenz \textit{et
al.} \cite{lorenz3} also report that the value of $T_{c}$ is degraded with
increasing lattice strain, a result confirmed by Serquis \textit{et al.}
\cite{serquis}. \ $T_{c}$ in MgB$_{2}$ has also been found to be lowered
following mechanical milling \cite{gao3} and after irradiation by fast
neutrons \cite{neutrons}. \ In contrast to the work of Lorenz \textit{et al.}
\cite{lorenz3} and Serquis \textit{et al.} \cite{serquis}, Hinks \textit{et
al.} \cite{hinks} have recently reported that accidental impurity doping can
have an effect on $T_{c}$ much larger than that of lattice strain.
\ Additionally, they show that grain-interaction stresses can significantly
alter the lattice parameters of MgB$_{2}$ depending on the impurity phases
present with the MgB$_{2}$. \ For example, the largest strains are seen for
samples where MgB$_{4}$ is present as an impurity phase, imparting to the bulk
sample the properties of an MgB$_{2}$/MgB$_{4}$ composite. One might
hypothesize that grain-interaction stresses could significantly alter the
response of MgB$_{2}$ grains to pressure in such samples, and, in this way,
modify the observed response of $T_{c}$ to pressure.

In contrast to the poor agreement between the large body of nonhydrostatic
$T_{c}(P)$ data on MgB$_{2},$ the agreement between the purely hydrostatic
He-gas data to 0.7 GPa and the nearly hydrostatic dense He data to 20 GPa (see
Fig. 4) is remarkably good for both single-crystalline and polycrystalline
samples. \ It is thus reasonable to assert that the intrinsic initial pressure
dependence for MgB$_{2}$ samples with the highest values of $T_{c}$ (39.1 K
for $^{11}$B) is given by $dT_{c}/dP\simeq$ -1.11(2) K/GPa, the $T_{c}(P)$
dependence to 32 GPa being given by the present data shown in Figs. 3 and 4.
\ We would now like to compare these experimental results with theoretical models.

\subsection{Comparison with Theory}

For most known superconductors, including MgB$_{2},$ $T_{c}$ is found to
decrease with pressure. \ In fact, $dT_{c}/dP$ is negative for \textit{all}
simple-metal superconductors (e.g. Pb, Al, Sn, and In \cite{eiling}) due to
pressure-induced lattice stiffening (higher phonon frequencies) which weakens
the electron-phonon coupling \cite{sp}. \ Rb$_{3}$C$_{60}$ and other
alkali-doped fullerenes are exceptions to this ``rule'': \ here the rapid
decrease in $T_{c}$ with pressure originates not from lattice stiffening but
rather from a sharp decrease in the electronic density of states $N(E_{f})$
\cite{diederichs}. \ What is the origin of the negative pressure dependence of
$T_{c}$ for MgB$_{2}$ - is it lattice stiffening, a decrease in the electronic
density of states, or something else?

Theoretical models are calculated in terms of the dependence of the relevant
properties on the lattice parameters or unit cell volume $V.$ \ From the
intrinsic initial slope $dT_{c}/dP\simeq$ -1.11 K/GPa for MgB$_{2}$ we can
calculate the logarithmic volume derivative
\begin{equation}
\frac{d\ln T_{c}}{d\ln V}=-\frac{B}{T_{c}(0)}\left(  \frac{dT_{c}}{dP}\right)
=+4.18,
\end{equation}
where we use $T_{c}(0)=$ 39.1 K from above and the bulk modulus $B=$ 147.2 GPa
from He-gas neutron diffraction studies on the same sample \cite{r22}.

We now convert the $T_{c}(P)$ data for polycrystalline MgB$_{2}$ in Fig. 4 to
$T_{c}(V)$ data using the Murnaghan equation of state \cite{murg}. \ In Fig. 7
we compare the resulting $T_{c}(V)$ dependences to that for the type ``B''
single crystal\ from Fig. 3; the agreement is excellent considering that
$T_{c}(0)$ for the single crystal is nearly 1 K lower than for the
polycrystalline sample.

We now attempt a fit to the data in Fig. 7 using the well known McMillan
formula \cite{mcmillan}%

\begin{equation}
T_{c}\simeq\frac{\left\langle \omega\right\rangle }{1.2}\exp\left\{
\frac{-1.04(1+\lambda)}{\lambda-\mu^{\ast}(1+0.62\lambda)}\right\}  ,
\end{equation}
which is valid for strong coupling ($\lambda\lesssim1.5)$ and connects the
value of $T_{c}$ with the electron-phonon coupling parameter $\lambda,$ an
average phonon frequency $\left\langle \omega\right\rangle ,$ and the Coulomb
repulsion $\mu^{\ast}$. \ We should not expect too much from this fit since
the McMillan formula is a solution of the isotropic Eliashberg equations and
thus ignores the strong anisotropies in the vibrational, electronic, and
superconducting properties of MgB$_{2}.$ \ In the following we draw heavily on
the detailed analysis of Chen \textit{et al. }\cite{chen2} who consider the
effect of pressure on the three relevant parameters in Eq. (2) $\left\langle
\omega\right\rangle ,$ $\lambda,$ and $\mu^{\ast},$ .

Taking the logarithmic volume derivative of both sides of Eq. (2), we obtain
the simple relation
\begin{equation}
\frac{d\ln T_{c}}{d\ln V}=-\gamma-\Delta_{1}\left\{  \frac{\partial\ln
\mu^{\ast}}{\partial\ln V}\right\}  +\Delta_{2}\left\{  \frac{\partial\ln\eta
}{\partial\ln V}+2\gamma\right\}  ,
\end{equation}
where $\gamma\equiv-\partial\ln\left\langle \omega\right\rangle /\partial\ln
V$ is the Gr\"{u}neisen parameter, $\eta\equiv N(E_{f})\left\langle
I^{2}\right\rangle $ is the McMillan-Hopfield parameter given by the product
of the electronic density of states and the average squared electronic matrix
element, and the dimensionless prefactors are given by $\Delta_{1}%
=1.04\mu^{\ast}(1+\lambda)\left[  1+0.62\lambda\right]  /\left[  \lambda
-\mu^{\ast}(1+0.62\lambda)\right]  ^{2}$ and $\Delta_{2}=1.04\lambda\left[
1+0.38\mu^{\ast}\right]  /\left[  \lambda-\mu^{\ast}(1+0.62\lambda)\right]
^{2}.$ Both $\Delta_{1}$ and $\Delta_{2}$ are calculated using the values of
$\lambda$ and $\mu^{\ast}$ at ambient pressure.

Eq. (3) looks rather formidable, but has a very simple interpretation.
\ Fortunately, the first two terms on the right are usually small relative to
the third term, as we will see below, so that $d\ln T_{c}/d\ln V\approx
\Delta_{2}\{\partial\ln\eta/\partial\ln V+2\gamma\}$. \ Since $\Delta_{2}$ is
always positive, the sign of the logarithmic derivative $d\ln T_{c}/d\ln V$ is
determined by the relative magnitude of the two terms $\partial\ln
\eta/\partial\ln V$ and $2\gamma$. \ The first ``electronic'' term is negative
($\partial\ln\eta/\partial\ln V\approx-1 $ for simple metals (s,p electrons)
\cite{sp}, but may equal -3 to -4 for transition metals (d
electrons)\cite{r17''}), whereas the second ``lattice'' term is positive
(typically $2\gamma\approx3-5).$ \ Since in simple-metal superconductors, like
Al, In, Sn, and Pb, the lattice term dominates over the electronic term, the
sign of $d\ln T_{c}/d\ln V$ is the same as that of $\{\partial\ln\eta
/\partial\ln V+2\gamma\}$, namely positive; this accounts for the universal
decrease of $T_{c}$ with pressure due to lattice stiffening in simple metals.
\ In selected transition metals, the electronic term may become larger than
the lattice term, in which case $T_{c}$ would be expected to increase with
pressure, as observed in experiment \cite{r17''}.

We now apply the McMillan equation to the above results on MgB$_{2}$ using at
ambient pressure the logarithmically averaged phonon energy from inelastic
neutron studies \cite{takahashi} $\left\langle \omega\right\rangle =670$ K,
$T_{c}(0)\simeq39.1$ K, and $\mu^{\ast}=0.1$ \cite{andersen3}$,$ yielding
$\lambda\simeq0.898$, $\Delta_{1}=0.558$ and $\Delta_{2}=1.76.$ \ From the
expression derived by Chen \textit{et al. }\cite{chen2} for s,p metals
$\partial\ln\eta/\partial\ln V=-[\partial\ln N(E_{f})/\partial\ln V]-2/3$ and
the value $\partial\ln N(E_{f})/\partial\ln V=+0.46$ from Loa and Syassen
\cite{r16}, the dependence of the Hopfield parameter is estimated to be
$\partial\ln\eta/\partial\ln V=-1.13,$ a value reasonably close to the generic
value (-1) used above in the analysis for the simple-metal superconductors and
to the value (-0.81) obtained from first-principles electronic structure
calculations on MgB$_{2}$ by Medvedera \textit{et al.} \cite{ivan}. \ Chen
\textit{et al. }\cite{chen2} find that $\mu^{\ast}$ increases only weakly with
pressure at a rate $\phi\equiv\partial\ln\mu^{\ast}/\partial\ln V=$
-0.1$\gamma-$ 0.035; note that this derivative is very small ($\sim$ -0.3) so
that the second term on the right side of Eq. (3) is relatively unimportant.
\ We now have estimates of all quantities on the right side of Eq. (3) except
$\gamma$ which we use as a fit parameter. \ Setting the left side of Eq. (3)
equal to the experimental value +4.18 from Eq. (1), we find $\gamma=2.39$, in
reasonable agreement with the value $\gamma\approx2.9$ from Raman spectroscopy
studies \cite{goncharov} or $\gamma\approx2.3$ from \textit{ab initio
}electronic structure calculations on MgB$_{2}$ \cite{roundy}. \ Note that for
the present ``type B'' crystal $T_{c}(0)\simeq38.24$ K and $dT_{c}%
/dP\simeq-1.10$ K/GPa which gives $d\ln T_{c}/d\ln V=+4.23,$ $\lambda
\simeq0.887$, and $\gamma=2.36.$

Eq. (3) is only valid for small changes in the parameters, i.e. for
experiments to a few GPa pressure where the change in unit cell volume is only
a few percent. \ To attempt to fit the very high pressure $T_{c}(V)$ data in
Figs. 3 and 7, we need to use the full McMillan equation and insert explicitly
the change in the parameters with relative volume. As suggested by Chen
\textit{et al.} \cite{chen2}, we set
\begin{equation}
\left\langle \omega\right\rangle =\left\langle \omega\right\rangle
_{0}(V/V_{0})^{-\gamma},\text{\ }\lambda=\lambda_{0}(V/V_{0})^{\varphi},\text{
and }\mu^{\ast}=\mu_{0}^{\ast}(V/V_{0})^{\phi},
\end{equation}
where $\varphi\equiv\{\partial\ln\eta/\partial\ln V+2\gamma\}.$ \ Using the
values of the parameters for the sample with $T_{c}(0)=39.1$ K used above to
fit the initial pressure dependence ($\gamma\simeq2.4,$ $\varphi
=-1.13+2\times2.4=3.67,$ and $\phi=-0.1\times2.4-0.035=-0.275$ and inserting
the volume dependences from Eq. (4) into the McMillan equation, the lower
solid fit curve in Fig. 7 is obtained. \ Note that this curve clearly lies
below the data at higher pressures. \ A reasonably good fit to the data over
the pressure range to 32 GPa ($V/V_{o}=0.855$) is found for $\gamma=2.2$ (see
Fig. 7). \ As with the simple $s,p$ metal superconductors, $T_{c}$ in
MgB$_{2}$ appears to decrease under pressure due to lattice stiffening. \ Note
that within this approximation $T_{c}$ approaches 0 K asymptotically at very
high pressures. \ The fit curve for $\gamma=2.2$ is predicted to fall below 1
K for $V/V_{o}=$ 0.73 which corresponds to an applied pressure of 93 GPa.
\ The fact that the experimental data can be well fit by the McMillan formula
with reasonable values of the parameters lends support to the view that
MgB$_{2}$ is a BCS superconductor with moderately strong electron-phonon
coupling. \ However, the relatively small value of the fit parameter
$\gamma=2.2$ compared to experiment (2.9 \cite{goncharov}) is cause for concern.

As discussed in the Introduction, the binary compound MgB$_{2}$ is a quasi 2D
system with highly anisotropic electronic and lattice properties, including
multiple superconducting gaps. \ The above analysis of experimental data using
the McMillan formula, which represents a solution to the isotropic Eliashberg
equations, is a good first step but does not permit unequivocal conclusions.
\ What is needed is an extension of the solution of the fully anisotropic
Eliashberg equations \cite{r4} to reduced lattice parameters. \ The initial
dependences $dT_{c}/dP\simeq-1.11(2)$ K/GPa and $d\ln T_{c}/d\ln V\simeq$
+4.18, and the $T_{c}(P)$ and $T_{c}(V)$ data to 32 GPa in Figs. 3, 4, and 7
stand ready to provide a stringent test of such a calculation.

\vspace{0.35cm}\noindent\textbf{Acknowledgements.} \ The authors are grateful
to V.G. Tissen for providing the sample used in his original studies. \ Thanks
are due V.V. Struzhkin for providing the numerical data from his original
publication. \ Work at Washington University is supported by NSF grant
DMR-0101809 and that at the Argonne National Laboratory by the U.S. Department
of Energy, Office of Science, contract No. W-31-109-ENG-38. \ The work in
Japan was partially supported by the New Energy and Industrial Technology
Development Organization (NEDO) as Collaborative Research and Development of
Fundamental Technologies for Superconductivity Applications.


\section{Figure Captions}

\bigskip

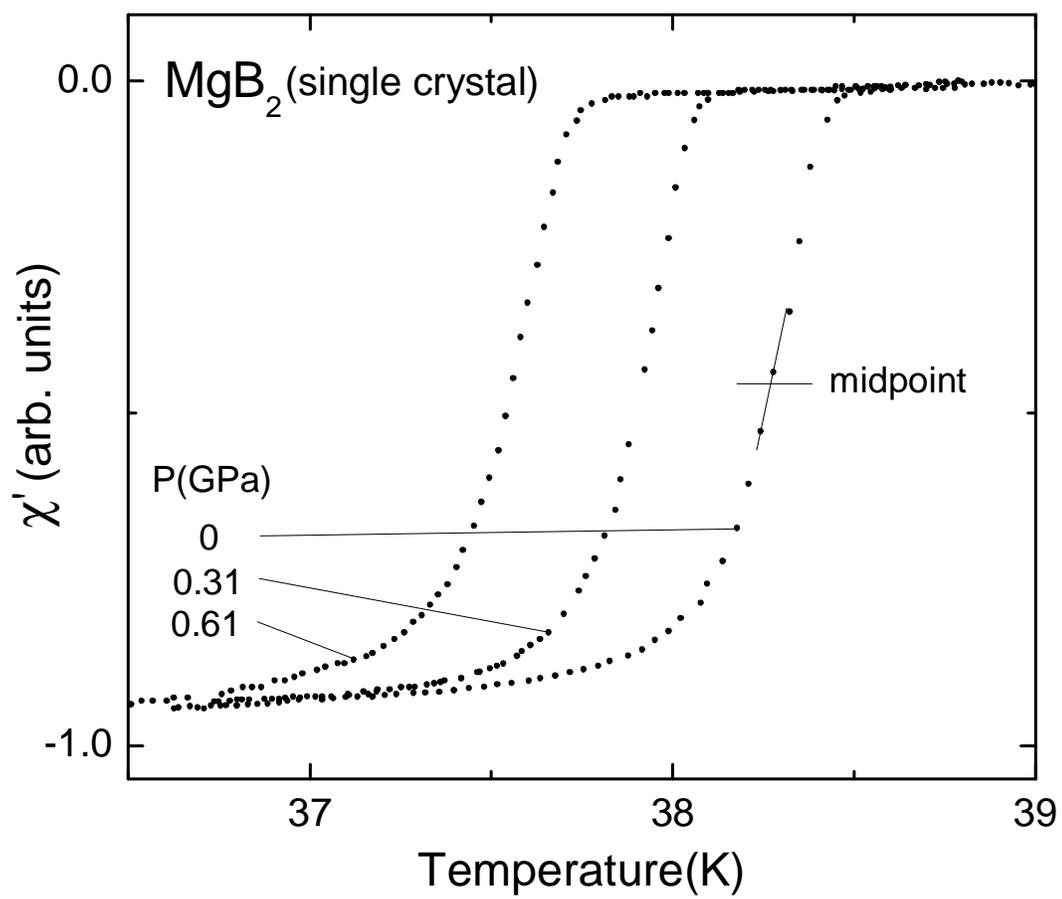
\noindent\textbf{Fig. 1. \ }Real part of the ac susceptibility $\chi^{\prime}$
(1023 Hz, 1 Oe (rms)) versus temperature for a ``type B'' MgB$_{2}$ single
crystal at three different pressures. \ The position of the superconducting
midpoint is indicated for ambient pressure data.

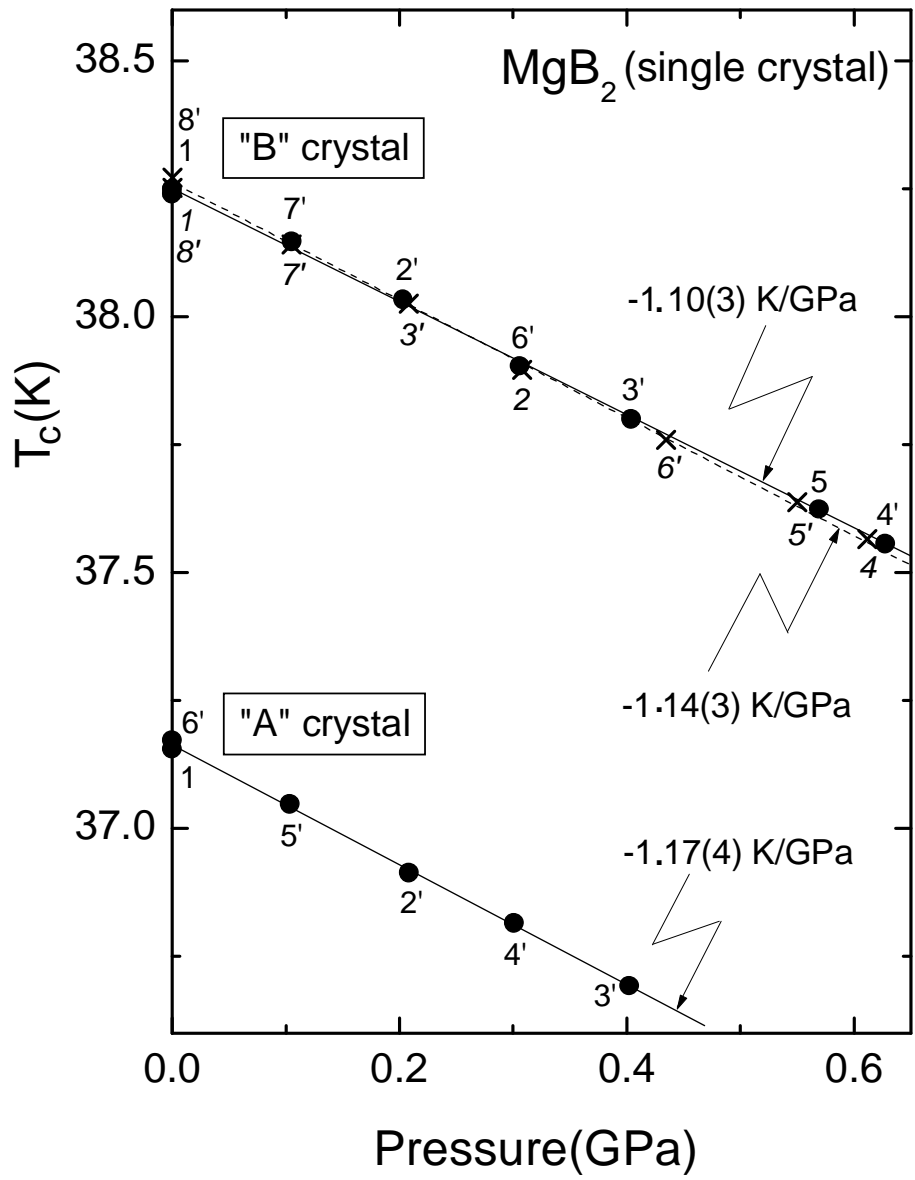
\noindent\textbf{Fig. 2. \ }Dependence of the superconducting transition
temperature of one ``type A'' ($\bullet$) and two ``type B'' ($\bullet,\times
$) MgB$_{2}$ single crystals on hydrostatic He-gas pressure. \ Numbers give
order of measurement. \ Pressure was normally changed at room temperature; for
data with primed numbers the pressure was changed at low temperatures ($\sim$
50 K). \ Solid and dashed lines are guides to the eye.

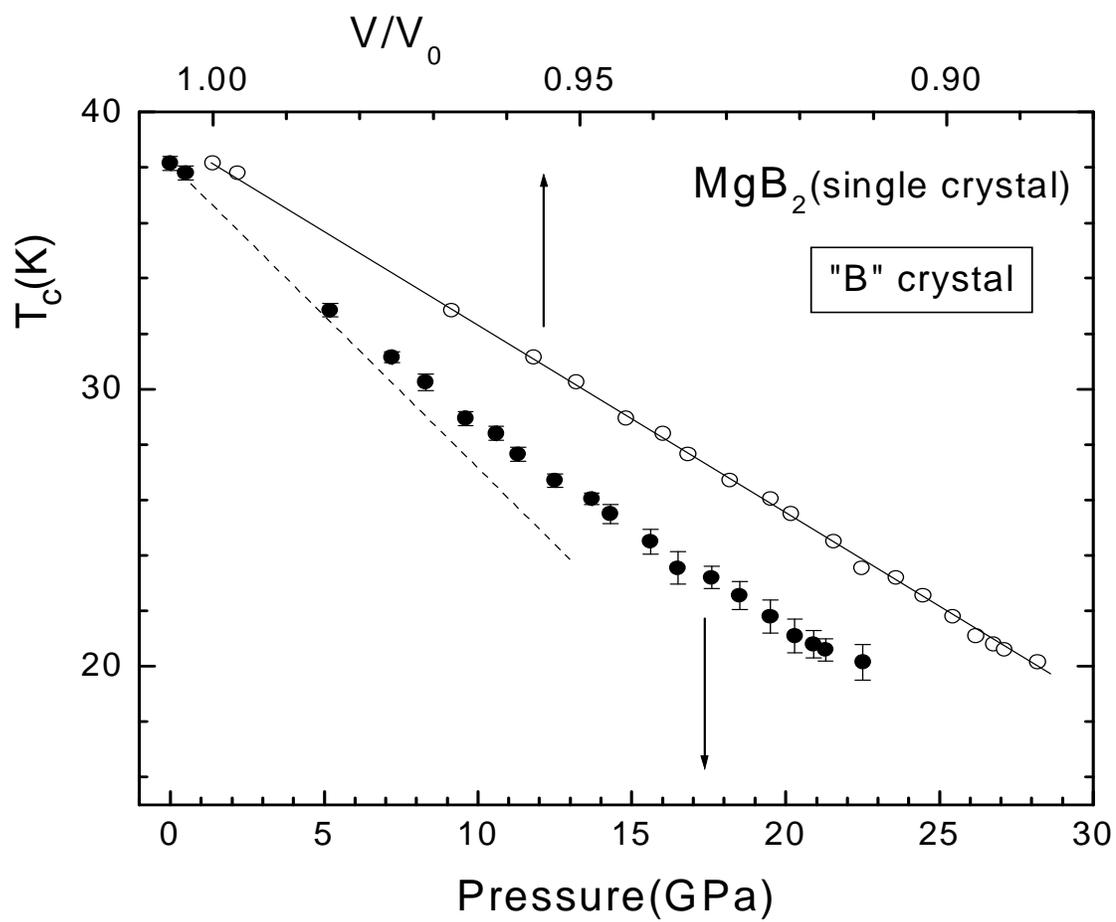
\noindent\textbf{Fig. 3. \ }Dependence of the superconducting transition
temperature of ``type B'' MgB$_{2}$ single crystal on nearly hydrostatic
pressure in a He-loaded DAC. \ All data taken in order of increasing pressure,
where pressure was only changed at ambient temperature. $\ T_{c}$ is
determined from the midpoint of the superconducting transition in the
temperature-dependent ac susceptibility $\chi^{\prime}$ (1003 Hz, 3 Oe (rms)).
\ Dashed line gives slope $dT_{c}/dP\simeq-1.10$ K/GPa of He-gas measurement
on same crystal (Fig. 2). \ The ``error bars'' give the temperatures of the
onset and end of the superconducting transition; note that the transition
broadens for $P\geq$ 14 GPa. \ $T_{c}$ versus relative volume $V/V_{o}$ is
also shown; straight solid line is guide to the eye.

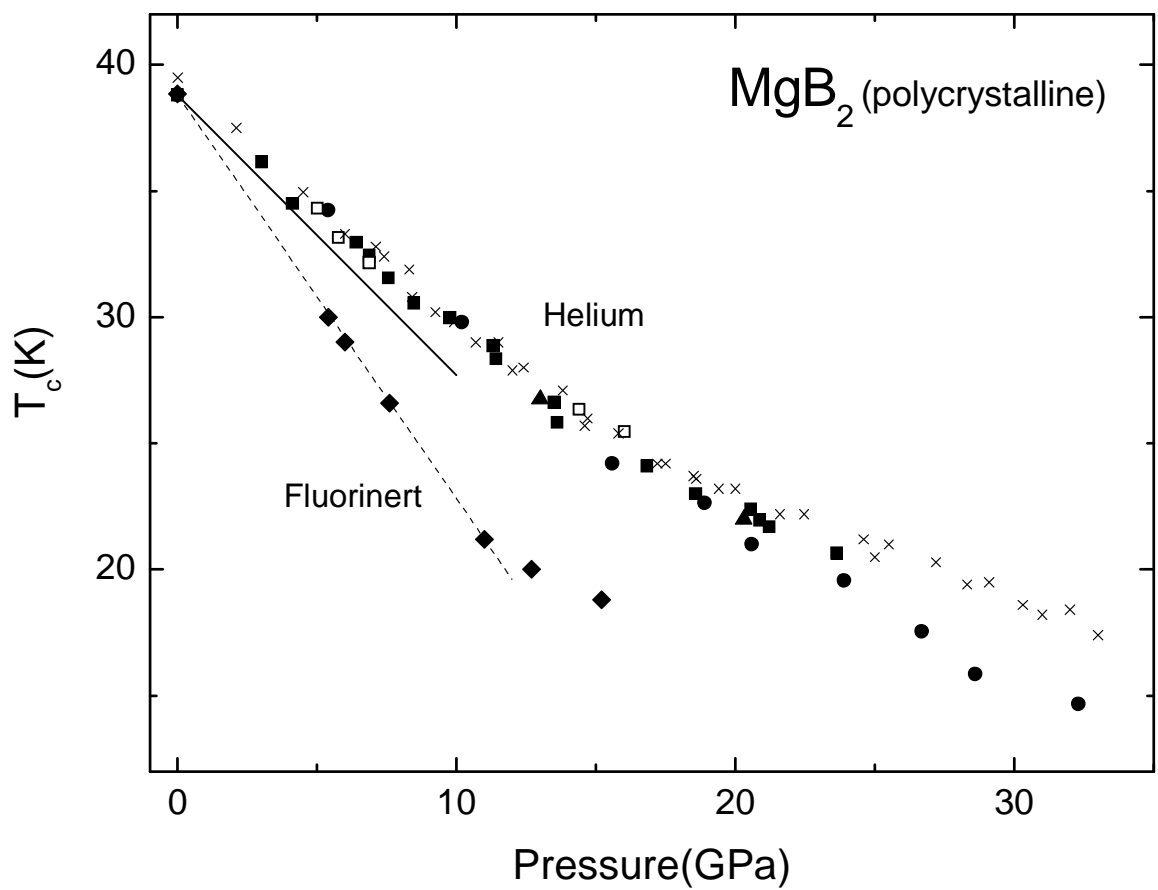
\noindent\textbf{Fig. 4. \ }Dependence of the superconducting transition
temperature of isotopically pure $^{11}$B polycrystalline MgB$_{2}$ on nearly
hydrostatic pressure in He-loaded DAC measurements: three different
experiments ($\bullet,\blacksquare,\square,\blacktriangle)$, closed symbols
(increasing pressure), open symbols (decreasing pressure)$;$ measurements from
Struzhkin \textit{et al. }\cite{r20} ($\times).$ \ Measurements in our DAC on
same sample with Fluorinert pressure medium ($\blacklozenge$) are also shown;
solid and dashed lines gives slopes (-1.11 K/GPa) and (-1.6 K/GPa) from He-gas
\cite{r15} and Fluorinert data, respectively, on same sample.

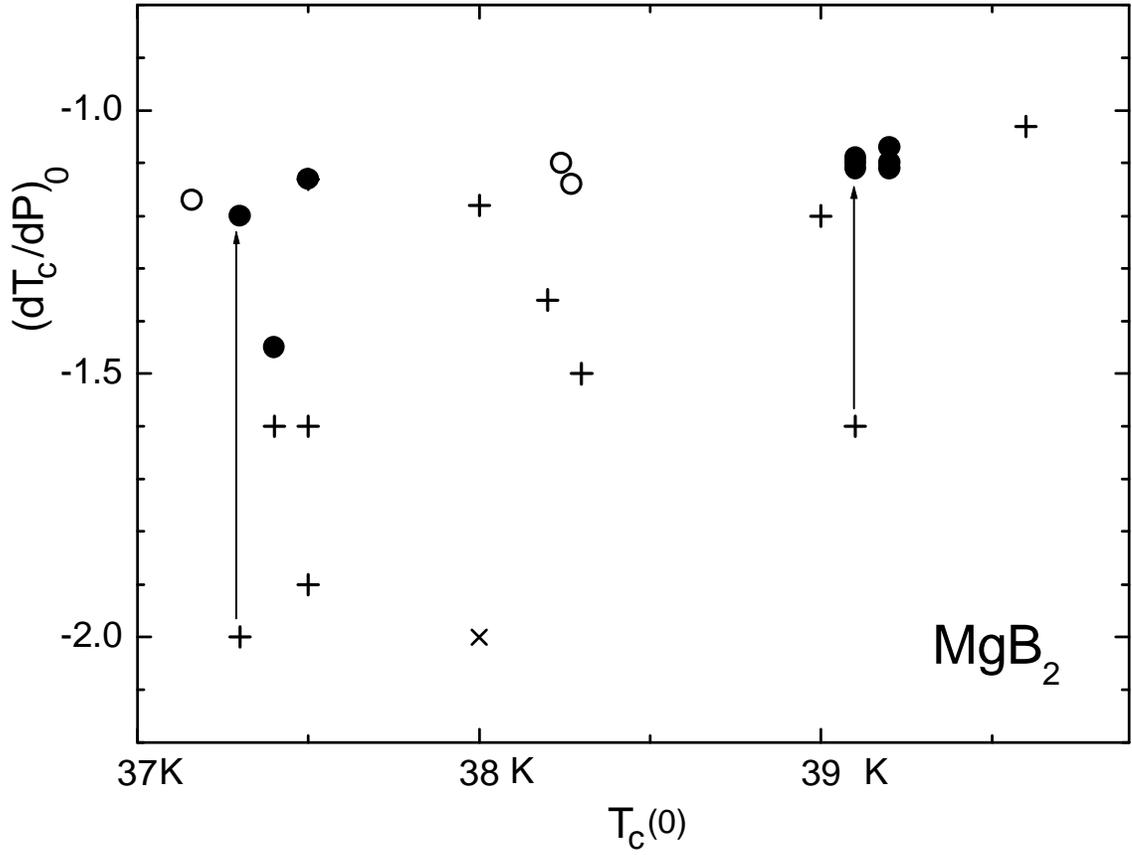
\noindent\textbf{Fig. 5. \ }Initial pressure derivative ($dT_{c}/dP)_{0}$
versus value of $T_{c}$ at ambient pressure$;$ data are taken directly from
the Table. \ Measurements with He ($\bullet,\circ$) and non-He ($+,\times$)
pressure media; both polycrystalline ($\bullet,+$) and single-crystalline
($\circ,\times$) samples are represented. \ Vertical arrows show change in
measured value of ($dT_{c}/dP)_{0}$ for a given sample upon changing to He
pressure medium (see text).

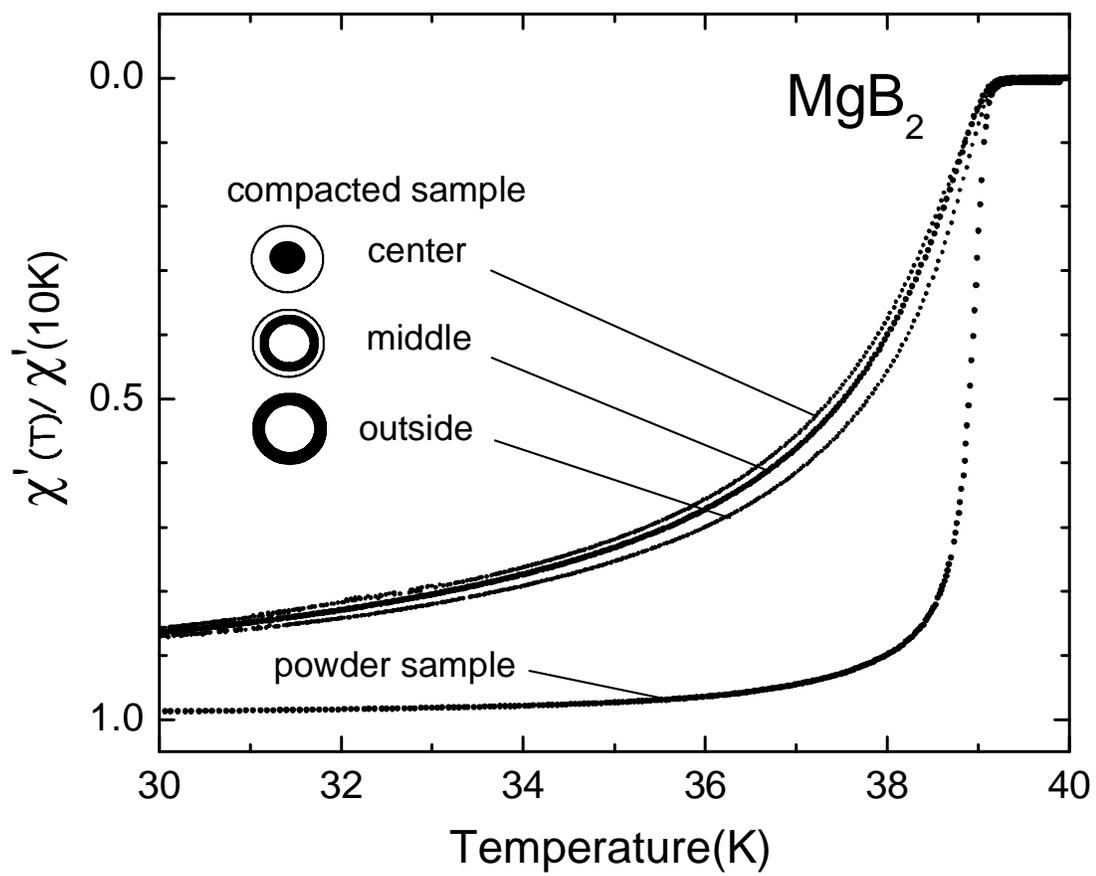
\noindent\textbf{Fig. 6. \ }Relative change in the real part of the ac
susceptibility $\chi^{\prime}$ of MgB$_{2}$ versus temperature for both loose
powder and flat, compacted samples. \ For the compacted disc, samples were
taken from the center, middle, and outside regions.

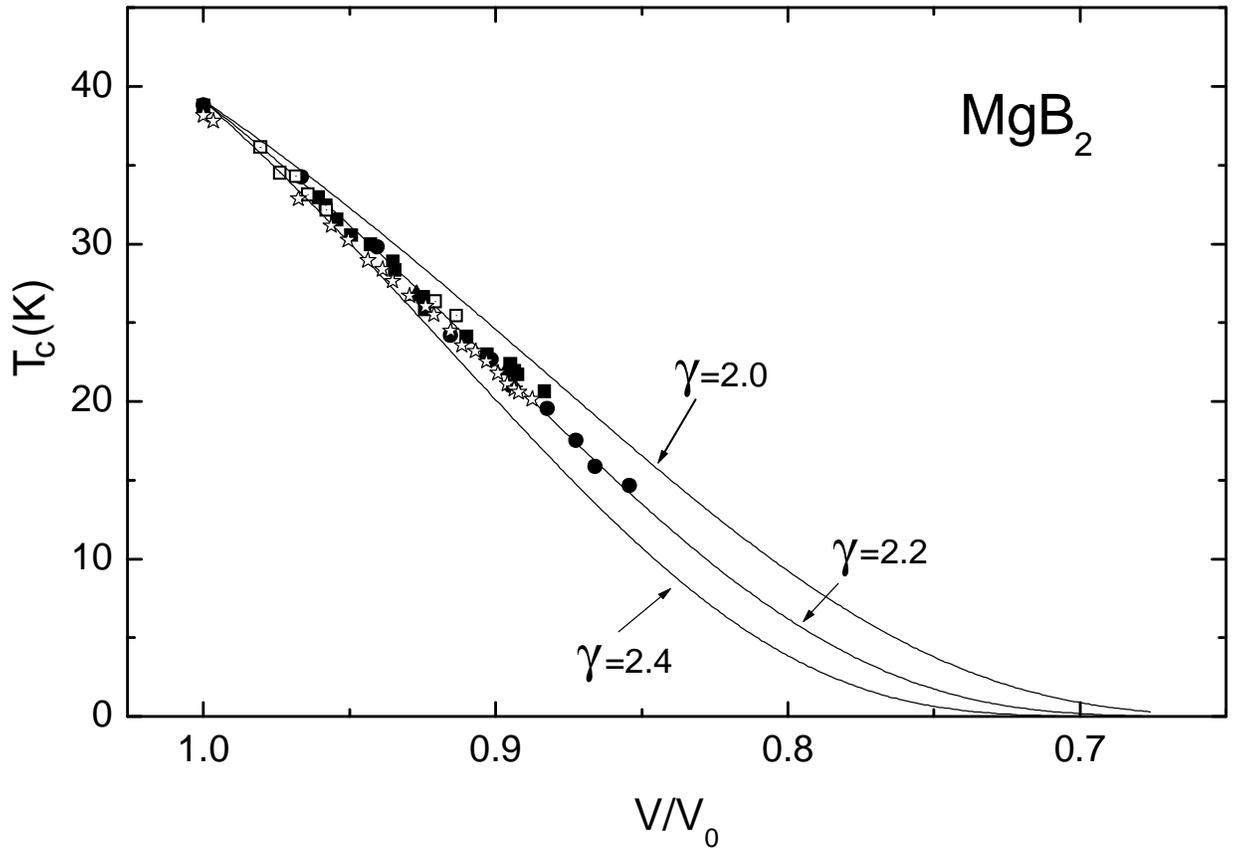
\noindent\textbf{Fig. 7. \ }$T_{c}$ values from present single-crystalline
(open stars) and polycrystalline ($\bullet,\blacksquare,\square,\blacktriangle
)$ data from Figs. 3 and 4 plotted versus relative volume. \ Solid lines are
calculated curves using McMillan's equation (Eq. 2) for three different values
of the Gr\"{u}neisen parameter $\gamma$ (see text).\newpage

\noindent\textbf{Table. \ }Summary of available high-pressure $T_{c}(P)$ data
on MgB$_{2}$ single-crystals (first 4 rows) and polycrystals (remaining rows).
\ $T_{c}$ values are at ambient pressure from the superconducting midpoint in
the ac susceptibility $\chi_{ac}$ and electrical resistivity $\rho$
measurements; Struzhkin \textit{et al.} \cite{r20} use a double-modulation
technique $\chi_{ac}^{\operatorname{mod}}$ which is believed to give the
superconducting onset. \ $(dT_{c}/dP)_{0}$ is the initial pressure derivative.
\ $P^{\max}$(GPa) is the maximum pressure reached in the experiment . \ Unless
otherwise specified, samples with the natural boron isotopic abundance
$^{10.81}$B are studied. \ Arrows indicate increasing ($\uparrow$) or
decreasing ($\downarrow$) pressure.\vspace{0.2cm}

\begin{center}%
\begin{tabular}
[c]{|l|l|l|l|l|l|}\hline
$%
\begin{array}
[c]{c}%
T_{c}(0)\\
\text{(K)}%
\end{array}
$ & $%
\begin{array}
[c]{c}%
(dT_{c}/dP)_{0}\\
\text{(K/GPa)}%
\end{array}
$ & $%
\begin{array}
[c]{c}%
P^{\max}\\
\text{(GPa)}%
\end{array}
$ & \textbf{measurement} & \textbf{pressure medium} & \textbf{reference}%
\\\hline\hline
38.24 & -1.10(3) & 0.63, 23 & $\chi_{ac}$, ``B'' crystal & helium & this
paper\\\hline
38.27 & -1.14(3) & 0.61 & $\chi_{ac}$, ``B'' crystal & helium & this
paper\\\hline
37.16 & -1.17(4) & 0.4 & $\chi_{ac}$, ``A'' crystal & helium & this
paper\\\hline
38.0 & -2.0 & 1.4 & $\rho$, crystal & Fluorinert & \cite{masui}\\\hline
&  &  &  &  & \\\hline
39.1 & -1.11(2) & 0.66 & $\chi_{ac}$, $^{11}$B & helium & \cite{r15}\\\hline
39.1 & -1.09(4) & 0.63 & $\chi_{ac}$, $^{11}$B & helium & \cite{r17}\\\hline
39.2 & -1.11(3) & 0.61 & $\chi_{ac}$, $^{11}$B & helium & \cite{r17}\\\hline
40.5 & -1.12(3) & 0.64 & $\chi_{ac}$, $^{10}$B & helium & \cite{r17}\\\hline
37.5 & -1.13 & 0.4 & $\chi_{ac}$ & helium & \cite{r19}\\\hline
39.2 & -1.07 & 0.84 & $\chi_{ac}$ & helium & \cite{r18}\\\hline
37.4 & -1.45 & 0.84 & $\chi_{ac}$ & helium & \cite{r18}\\\hline
37.3 & -1.2 & 0.6 & $\chi_{ac}$ & helium & \cite{tissen2}\\\hline
39.1 & -1.1 & 32.3 & $\chi_{ac}$, $^{11}$B & helium & this paper\\\hline
40.2 & -1.1 & 33 & $\chi_{ac}^{\operatorname{mod}}$, $^{11}$B & helium &
\cite{r20}\\\hline
39.2 & -1.1 & 44 & $\chi_{ac}^{\operatorname{mod}}$, $^{10}$B & helium &
\cite{r20}\\\hline
&  &  &  &  & \\\hline
39.1 & -1.6 & 15 & $\chi_{ac}^{\operatorname{mod}}$, $^{11}$B & Fluorinert &
this paper\\\hline
37.4 & -1.6 & 1.84 & $\chi_{ac}$ & Fluorinert & \cite{r13}\\\hline
37.3 & -2 & 28 & $\chi_{ac}$ & 4:1 methanol/ethanol & \cite{r6}\\\hline
38.2 & -1.36 & 1.46 & $\rho$ & 1:1 daphne/kerosene & \cite{choi}\\\hline
37.5 & -1.9 & 1.35 & $\rho$ & Fluorinert & \cite{r14}\\\hline
38.3 & -1.5(1) & 1.1 & $\chi_{dc}$ & kerosene/mineral oil & \cite{kazakov}%
\\\hline
39.6 & -1.03 & 9 & $\rho$ & Fluorinert & \cite{tang}\\\hline
38 & -1.18(6) & 0.8 & $\chi_{dc}$ & silicon oil & \cite{razavi}\\\hline
&  &  &  &  & \\\hline
37.5 & -1.6 (P $\uparrow$) & 7.6 & $\chi_{ac}$ & NaF & \cite{r19}\\\hline
37.5 & -1.13 (P $\downarrow$) & 7.6 & $\chi_{ac}$ & NaF & \cite{r19}\\\hline
39 & -1.20(9) & 11 & $\rho$ & steatite & \cite{razavi}\\\hline
$\sim$ 35 & -0.35 to -0.8 & 33 & $\rho$ & steatite & \cite{r12,bordet}\\\hline
\end{tabular}
\end{center}

\begin{thebibliography}{9}                                                                                                %

\bibitem {r0}J. Nagamatsu, N. Nakagawa, T. Muranaka, Y. Zenitani, J. Akimitsu,
Nature 410 (2001) 63.

\bibitem {r1}See other papers in this special issue; see also the review: \ C.
Buzea and T. Yamashita, Supercond. Sci. Technol. 14 (2001) R115.

\bibitem {r2}M.K. Wu, J.R. Ashburn, C.J. Torng, P.H. Hor, R.L. Meng, L. Gao,
Z.J. Huang, Y.Q. Wang, and C.W. Chu, Phys. Rev. Lett. 58 (1987) 908.

\bibitem {r3}J.G. Bednorz and K.A. M\"{u}ller, Z. Phys. B 64 (1986) 189.

\bibitem {r22}J.D. Jorgensen, D.G. Hinks, S. Short, Phys. Rev. B 63 (2001) 224522.

\bibitem {kobayashi1}K. Kobayashi, K. Yamamoto, J. Phys. Soc. Japan 70 (2001) 1861.

\bibitem {r24}X. Wan, J. Dong, H. Weng, D.Y. Xing, Phys. Rev. B 65 (2001) 012502.

\bibitem {hc21}M. Angst, R. Puzniak, A. Wisniewski, J. Jun, S.M. Kazakov, J.
Karpinski, J. Roos, H. Keller, Phys. Rev. Lett. 88 (2002) 167004.

\bibitem {hc22}Yu. Eltsev. S. Lee, K. Nakao, N. Chikumoto, S. Tajima, N.
Koshizuka, M. Murakami, Physica C (2002) in press.

\bibitem {gaps}See, for example: \ P. Szabo, P. Samuely, J. Kamar\'{\i}k, T.
Klein, J. Marcus, D. Fruchart, S. Miraglia, C. Marcenat, A.G.M. Jansen, Phys.
Rev. Lett. 87 (2001) 137005; F. Bouquet, R.A. Fisher, N.E. Phillips, D.G.
Hinks, J.D. Jorgensen, Phys. Rev. Lett. 87 (2001) 47001; H. Schmidt, J.F.
Zasadzinski, K.E. Gray, D.G. Hinks, preprint cond-mat/0112144.

\bibitem {theory1}A.Y. Liu, I.I. Mazin, J. Kortus, Phys. Rev. Lett. 87 (2001) 087005.

\bibitem {theory2}H.J. Choi, D. Roundy, H. Sun, M.L. Cohen, S.G. Louie,
preprint cond-mat/0111183.

\bibitem {r4}H.J. Choi, D. Roundy, H. Sun, M.L. Cohen, and S.G. Louie, Phys.
Rev. B 66 (2002) 020513; \textit{ibid.}, Nature 418 (2002) 758.

\bibitem {gonnelli}R.S. Gonnelli, D. Daghero, G.A. Ummarino, V.A. Stepanov, J.
Jun, S.M. Kazakov, J. Karpinski, preprint in cond-mat/0208060.

\bibitem {sulfur}V.V. Struzhkin, R.J. Hemley, H.K. Mao, Y. Timofeev, Nature
390 (1997) 382.

\bibitem {oxygen}K. Shimizu, K. Suhara, M. Ikumo, M.I. Eremets, K. Amaya,
Nature 393 (1998) 767.

\bibitem {hpht}See, for example, Y. Takano, H. Takeya, H. Fujii, H. Kumakura,
T. Hatano, K. Togano, H. Kito, H. Ihara, Appl. Phys. Lett 78 (2001) 2914; C.U.
Jung, M.-S. Park, W.N. Kang, M.-S. Kim, K.H.P. Kim, S.Y. Lee, S.-I. Lee, Appl.
Phys. Lett. 78 (2001) 4157; P. Toulemonde, N. Musolino, R. Flukiger, preprint
cond-mat/0207033; T.A. Prikhna, W. Gawalek, A.B. Surzhenko, N.V. Sergienko,
V.E. Moshchil, T. Habisreuther, V.S. Melnikov, S.N. Dub, P.A. Nagorny, M.
Wendt, Ya.M. Savchuk, D. Litzkendorf, J. Dellith, S. Kracunovska, Ch. Schmidt,
preprint cond-mat/0109216.

\bibitem {lee}S. Lee, T. Masui, H. Mori, Yu. Eltsev, A. Yamamoto, S. Tajima,
preprint cond-mat/0207247.

\bibitem {r5}J. Karpinski, M. Angst, J. Jun, S.M. Kazakov, R. Puzniak, A.
Wisniewski, J. Roos, H. Keller, A. Perucchi, L. Degiorgi, M. Eskildsen, P.
Bordet, L. Vinnikov, A. Mironov, preprint cond-mat/0207263.

\bibitem {jung}C.U. Jung, J.-H. Choi, P. Chowdhury, K.H.P. Kim, M.-S. Park,
H.-J. Kim, J.Y. Kim, Z. Du, M.-S. Kim, W.N. Kang, S.I. Lee, G.Y. Sung, J.Y.
Lee, preprint cond-mat/0105330.

\bibitem {brand}N.B. Brandt, N.I. Ginzburg, Sci. Amer. 224 (1971) 83.

\bibitem {jerome}D. Jerome, H.J. Schulz, Adv. Phys. 51 (2002) 293;
\textit{ibid.}, Adv. Phys. 31 (1982) 299.

\bibitem {sp}J.S. Schilling and S. Klotz, in: \ \textit{Physical Properties of
High Temperature Superconductors, }Vol. III, ed. D.M. Ginsberg (World
Scientifc, Singapore, 1992) p. 59.

\bibitem {r12}M. Monteverde, M. N\'{u}\~{n}ez-Regueiro, N. Rogado, K.A. Regan,
M.A. Hayward, T. He, S.M. Loureiro, R.J. Cava, Science 292 (2001) 75.

\bibitem {r13}B. Lorenz, R.L. Meng, C.W. Chu, Phys. Rev. B 64 (2001) 012507.

\bibitem {r14}E. Saito, T. Taknenobu, T. Ito, Y. Iwasa, K. Prassides, T.
Arima, J. Phys. Condens. Matter 13 (2001) L267.

\bibitem {r15}T. Tomita, J.J. Hamlin, J.S. Schilling, D.G. Hinks, J.D.
Jorgensen, Phys. Rev. B 64 (2001) 092505.

\bibitem {r17}S. Deemyad, J.S. Schilling, J.D. Jorgensen, D.G. Hinks, Physica
C 361 (2001) 227.

\bibitem {r18}B. Lorenz, R.L. Meng, C.W. Chu, preprint cond-mat/0104303.

\bibitem {r19}S.I. Schlachter, W.H. Fietz, K. Grube, W. Goldacker, Advances in
Cryogenic Engineering: \ Proceedings of the International Cryogenic Materials
Conference - ICMC, Vol. 48 (2002) 809.

\bibitem {razavi}F.S. Razavi, S.K. Bose, H. Ploczek, Physica C 366 (2002) 73.

\bibitem {choi}E.S. Choi, W. Kang, J.Y. Kim, M.-S. Park, C.U. Jung, H.-J. Kim,
and S.-I. Lee, preprint cond-mat/0104454.

\bibitem {kazakov}S.M. Kazakov, M. Angst, J. Karpinski, I.M. Fita, R. Puzniak,
Solid State Commun. 119 (2001) 1.

\bibitem {masui}T. Masui, K. Yoshida, S. Lee, A. Yamamoto, S. Tajima, Phys.
Rev. B 65 (2002) 214513.

\bibitem {eremets}M.I. Eremets, \textit{High Pressure Experimental Methods
}(Oxford University Press, Oxford, 1996).

\bibitem {r9}J.E. Schirber, Cryogenics 10 (1970) 418.

\bibitem {r11}J. S. Schilling, J. Diederichs, S. Klotz, R. Sieburger, in:
\ \textit{Magnetic Susceptibility of Superconductors and Other Spin Systems},
edited by R. A. Hein, T. L. Francavilla, D. H. Liebenberg (Plenum Press, New
York, 1991), p. 107.

\bibitem {hinks}D.G. Hinks, J.D. Jorgensen, H. Zheng, S. Short, preprint cond-mat/0207161.

\bibitem {midpoint}We use the midpoint of the superconducting transition in
$\chi^{\prime}$ to define $T_{c}$ rather than the onset. \ Since the
transition broadens somewhat at higher pressures due to pressure inhomogeneity
and/or strain effects from the solid He pressure medium, the midpoint would be
expected to give a temperature closer to the average in the sample than the onset.

\bibitem {crystal}The small ``type B'' crystal (approximately $0.09\times
0.09\times0.03$ mm$^{3})$ used in the DAC experiment was actually cut off from
the larger crystal used to obtain the He-gas data ($\bullet$) in Fig. 2.

\bibitem {murg}We use the Murnaghan equation-of-state $V/V_{o}=[1+B^{\prime
}P/B]^{-1/B^{\prime}}$ with the value $B=147.2$ GPa from Ref. \cite{r22} and
the canonical value $B^{\prime}\equiv dB/dP=4$ supported by the $V(P)$ data to
40 GPa \cite{bordet} and a recent calculation \cite{r16}.

\bibitem {r21}The $T_{c}(P)$ data to 24 GPa are taken from Ref. \cite{r17}
after correcting the pressure values for a spectrometer miscalibration.

\bibitem {r20}V.V. Struzhkin, A.F. Goncharov, R.J. Hemley, H-k. Mao, G.
Lapertot, S.L. Bud'ko, P.C. Canfield, preprint cond-mat/0106576; see also
paper by Goncharov and Struzhkin in this special edition of Physica C; for
$T_{c}(P)$ data to 15 GPa see, A.F. Goncharov, V.V. Struzhkin, E. Gregoryanz,
H.K. Mao, R.J. Hemley, G. Lapertot, S.L. Bud'ko, P.C. Canfield, I.I. Mazin in:
\ ``Studies of High Temperature Superconductors'', Vol. 38, edited by A.V.
Narlikar (Nova Science Publishers, N.Y., 2001).

\bibitem {tang}J. Tang, L-C. Qin, A. Matsushita, Y. Takano, K. Togano, H.
Kito, H. Ihara, Phys. Rev. B 64 (2001) 132509.

\bibitem {r6}V.G. Tissen, M.V. Nefedova, N.N. Kolesnikov, M.P. Kulakov,
Physica C 363 (2001) 194.

\bibitem {bordet}P. Bordet, M. Mezouar, M. N\'{u}\~{n}ez-Regueiro, M.
Monteverde, M.D. N\'{u}\~{n}ez-Regueiro, N. Rogado, K.A. Regan, M.A. Hayward,
T. He, S.M. Loureiro, R.J. Cava, Phys. Rev. B 64 (2001) 172502.

\bibitem {lorenz3}B. Lorenz, Y.Y. Xue, R.L. Meng, C.W. Chu, preprint cond-mat/0110125.

\bibitem {tissen2}S. Deemyad, J.S. Schilling, V.G. Tissen, N.N. Koleshnikov,
M.P. Kulakov (unpublished).

\bibitem {serquis}A. Serquis, Y.T. Zhu, E.J. Peterson, J.Y. Coulter, D.E.
Peterson, F.M. Mueller, Appl. Phys. Lett. 79 (2001) 4399.

\bibitem {gao3}Y.D. Gao, J. Ding, G.V.S. Rao, B.V.R. Chowdar, W.X. Sun, Z.X.
Shen, Phys. Stat. Sol. (a) 191 (2002) 548.

\bibitem {neutrons}Y. Wang, F. Bouquet, I. Sheikin, P. Toulemonde, B. Revaz,
M. Eisterer, H.W. Weber, J. Hinderer, A. Junod, preprint cond-mat/0208169;
A.E. Karkin, V.I. Voronin, T.V. Dyachkova, A.P. Tyutyunnik, V.G. Zubkov, Yu.
G. Zainulin, B.N. Goshchitskii, preprint cond-mat/0103322.

\bibitem {eiling}See, for example, A. Eiling, J.S. Schilling, J. Phys. F 11
(1981) 623.

\bibitem {diederichs}J. Diederichs, A.K. Gangopadhyay, J.S. Schilling, Phys.
Rev. B 54 (1996) R9662.

\bibitem {mcmillan}W. L. McMillan, Phys. Rev. 167 (1968) 331.

\bibitem {chen2}X.J. Chen, H. Zhang, H.-U. Habermeier, Phys. Rev. B 65 (2002) 144514.

\bibitem {r17''}J.J. Hopfield, Physica (Amsterdam) 55 (1971) 41.

\bibitem {takahashi}T. Takahashi, T. Sato, S. Souma, T. Muranaka, J. Akimitsu,
Phys. Rev. Letters 86 (2001) 4915.

\bibitem {andersen3}Y. Kong, O.V. Dolgov, O. Jepsen, O.K. Andersen, Phys. Rev.
B 64 (2001) 020501.

\bibitem {r16}I. Loa and K. Syassen, Solid State Commun. 118 (2001) 279.

\bibitem {ivan}N.I. Medvedera, A.L. Ivanovskii, J.E. Medvedeva, A.J. Freeman,
D.L. Novokov, Phys. Rev. B 65 (2002) 052501.

\bibitem {goncharov}A.F. Goncharov, V.V. Struzhkin, E. Gregoryanz, J. Hu, R.J.
Hemley, Ho-k. Mao, G. Lapertot, S.L. Bud'ko, P.C. Canfield, Phys. Rev. B 64
(2001) 100509R.

\bibitem {roundy}D. Roundy, H. J. Choi, H. Sun, S. G. Louie, M. L. Cohen
(unpublished)\vspace{0.35cm}
\end{thebibliography}
\end{document}